\documentclass[%
reprint,
superscriptaddress,
 amsmath,amssymb,
 aps,
 pra,
]{revtex4-1}

\usepackage[utf8]{inputenc}
\usepackage{amssymb}
\usepackage{url}
\usepackage{float}
\usepackage{graphicx}
\usepackage{xcolor}
\usepackage{array}
\usepackage{tabularx}
\usepackage{booktabs}
\usepackage[colorlinks=true, citecolor=blue,urlcolor=blue,linkcolor=blue,filecolor=black]{hyperref}
\usepackage{cleveref}

\begin{document}

\preprint{AIP/123-QED}

\title[]{
Quantum Key Distribution using Deterministic Single-Photon Sources over a Field-Installed Fibre Link}

\author{Mujtaba Zahidy}%
\thanks{These authors contributed equally to this work.}
\affiliation{ 
Centre of Excellence for Silicon Photonics for Optical Communications (SPOC), Department of Electrical and Photonics Engineering, Technical University of Denmark, Kgs. Lyngby, Denmark
}%

\author{Mikkel T. Mikkelsen}%
\thanks{These authors contributed equally to this work.}
\affiliation{ 
Center for Hybrid Quantum Networks (Hy-Q), Niels Bohr Institute, University of Copenhagen, Blegdamsvej 17, Copenhagen, 2100, Denmark
}%

\author{Ronny M\"uller}%
\affiliation{ 
Centre of Excellence for Silicon Photonics for Optical Communications (SPOC), Department of Electrical and Photonics Engineering, Technical University of Denmark, Kgs. Lyngby, Denmark
}%

\author{Beatrice Da Lio}%
\affiliation{ 
Center for Hybrid Quantum Networks (Hy-Q), Niels Bohr Institute, University of Copenhagen, Blegdamsvej 17, Copenhagen, 2100, Denmark
}%

\author{Martin Krehbiel}%
\affiliation{ 
Center for Hybrid Quantum Networks (Hy-Q), Niels Bohr Institute, University of Copenhagen, Blegdamsvej 17, Copenhagen, 2100, Denmark
}%

\author{Ying Wang}%
\affiliation{ 
Center for Hybrid Quantum Networks (Hy-Q), Niels Bohr Institute, University of Copenhagen, Blegdamsvej 17, Copenhagen, 2100, Denmark
}%

\author{Nikolai Bart}
\affiliation{
Ruhr-Universit{\"a}t Bochum, Lehrstuhl f{\"u}r Angewandte Festk{\"o}perphysik, Universit{\"a}tsstrasse 150, Bochum, D-44780, Germany
}

\author{Andreas D. Wieck}
\affiliation{
Ruhr-Universit{\"a}t Bochum, Lehrstuhl f{\"u}r Angewandte Festk{\"o}perphysik, Universit{\"a}tsstrasse 150, Bochum, D-44780, Germany
}

\author{Arne Ludwig}
\affiliation{
Ruhr-Universit{\"a}t Bochum, Lehrstuhl f{\"u}r Angewandte Festk{\"o}perphysik, Universit{\"a}tsstrasse 150, Bochum, D-44780, Germany
}

\author{Michael Galili}%
\affiliation{ 
Centre of Excellence for Silicon Photonics for Optical Communications (SPOC), Department of Electrical and Photonics Engineering, Technical University of Denmark, Kgs. Lyngby, Denmark
}%

\author{S{\o}ren Forchhammer}%
\affiliation{ 
Centre of Excellence for Silicon Photonics for Optical Communications (SPOC), Department of Electrical and Photonics Engineering, Technical University of Denmark, Kgs. Lyngby, Denmark
}%

\author{Peter Lodahl}%
\affiliation{ 
Center for Hybrid Quantum Networks (Hy-Q), Niels Bohr Institute, University of Copenhagen, Blegdamsvej 17, Copenhagen, 2100, Denmark
}%

\author{Leif K. Oxenl{\o}we}%
\affiliation{ 
Centre of Excellence for Silicon Photonics for Optical Communications (SPOC), Department of Electrical and Photonics Engineering, Technical University of Denmark, Kgs. Lyngby, Denmark
}%

\author{Davide Bacco}%
\affiliation{ 
Department of Physics and Astronomy, University of Florence, Via Nello Carrara, Firenze, 50019, Italy
}%

\author{Leonardo Midolo}%
\affiliation{ 
Center for Hybrid Quantum Networks (Hy-Q), Niels Bohr Institute, University of Copenhagen, Blegdamsvej 17, Copenhagen, 2100, Denmark
}%

%\author*[1]{\fnm{Mujtaba} \sur{Zahidy}}%\email{muzah@fotonik.dtu.dk}

%\author[2]{\fnm{Mikkel T.} \sur{Mikkelsen}}%\email{iiauthor@gmail.com}

%\author[1]{\fnm{Ronny} \sur{M\"uller}}%\email{iiiauthor@gmail.com}

%\author[2]{\fnm{Beatrice} \sur{Da Lio}}%\email{iiiauthor@gmail.com}

%\author[2]{\fnm{Martin} \sur{Krehbiel}}%\email{iiiauthor@gmail.com}

%\author[2]{\fnm{Ying} \sur{Wang}}%\email{iiiauthor@gmail.com}

%\author[1]{\fnm{Michael} \sur{Galili}}%\email{iiiauthor@gmail.com}

%\author[1]{\fnm{S{\o}ren} \sur{Forchhammer}}%\email{iiiauthor@gmail.com}

%\author[2]{\fnm{Peter} \sur{Lodahl}}%\email{iiiauthor@gmail.com}

%\author[1]{\fnm{Leif K.} \sur{Oxenl{\o}we}}%\email{iiiauthor@gmail.com}

%\author[1,3]{\fnm{Davide} \sur{Bacco}}\email{dabac@fotonik.dtu.dk}

%\author[2]{\fnm{Leonardo} \sur{Midolo}}\email{midolo@nbi.ku.dk}

%\affil*[1]{\orgdiv{Department of Electrical and Photonics Engineering}, \orgname{Technical University of Denmark}, \orgaddress{\street{Ørsteds Pl.}, \city{Kgs. Lyngby}, \postcode{2800}, \country{Denmark}}}

%\affil[2]{\orgdiv{Center for Hybrid Quantum Networks (Hy-Q), Niels Bohr Institute}, \orgname{University of Copenhagen}, \orgaddress{\street{Blegdamsvej 17}, \city{Copenhagen}, \postcode{2100}, \country{Denmark}}}

%\affil[3]{\orgdiv{Department of Physics and Astronomy}, \orgname{University of Florence}, \orgaddress{\street{Via Nello Carrara}, \city{Firenze}, \postcode{50019}, \country{Italy}}}

\begin{abstract}
Quantum-dot-based single-photon sources are key assets for quantum information technology, supplying on-demand scalable quantum resources for computing and communication. However, long-lasting issues such as limited long-term stability and source brightness have traditionally impeded their adoption in real-world applications. Here, we realize a quantum key distribution field trial using true single photons across an 18-km-long dark fibre, located in the Copenhagen metropolitan area, using an optimized, state-of-the-art, quantum-dot single-photon source frequency-converted to the telecom wavelength. A secret key generation rate of $>$2 kbits/s realized over a 9.6 dB channel loss is achieved with a polarization-encoded BB84 scheme, showing remarkable stability for more than 24 hours of continuous operation. Our results highlight the maturity of deterministic single-photon source technology while paving the way for advanced single-photon-based communication protocols, including fully device-independent quantum key distribution, towards the goal of a quantum internet.
\end{abstract}

\maketitle

\section{Introduction}\label{sec1}
With the fast-growing developments of photonic-based quantum information technology, the demand for a reliable and deployable deterministic source of single photons has risen to new heights. In recent years, semiconductor quantum dots (QDs) embedded in photonic nanostructures have drawn significant attention by providing a robust and near-deterministic source of single photons \cite{Uppu2021}. The ability to generate indistinguishable photons on demand and couple them into optical fibers with high efficiency opens new avenues for the realization of a quantum internet \cite{Kimble_2008,lu2021quantum}, where photons will allow the secure exchange of secret cryptographic keys via quantum key distribution (QKD) or more advanced functionalities enabled by distributing quantum information via teleportation or entanglement swapping. To achieve this goal, it is essential that single-photon sources achieve sufficient quality and technological readiness to be operated in the field where virtually no control is available over sources of noise or loss of network infrastructure. Examples of quantum communication field trials using quantum emitters such as color centers in diamond \cite{hensen2015loophole} or semiconductor QDs \cite{basso2021quantum} are to date limited to short distances within university campuses over dedicated fibres or in free space.

\begin{figure}[ht!]
   \centering
    \includegraphics[width=\linewidth]{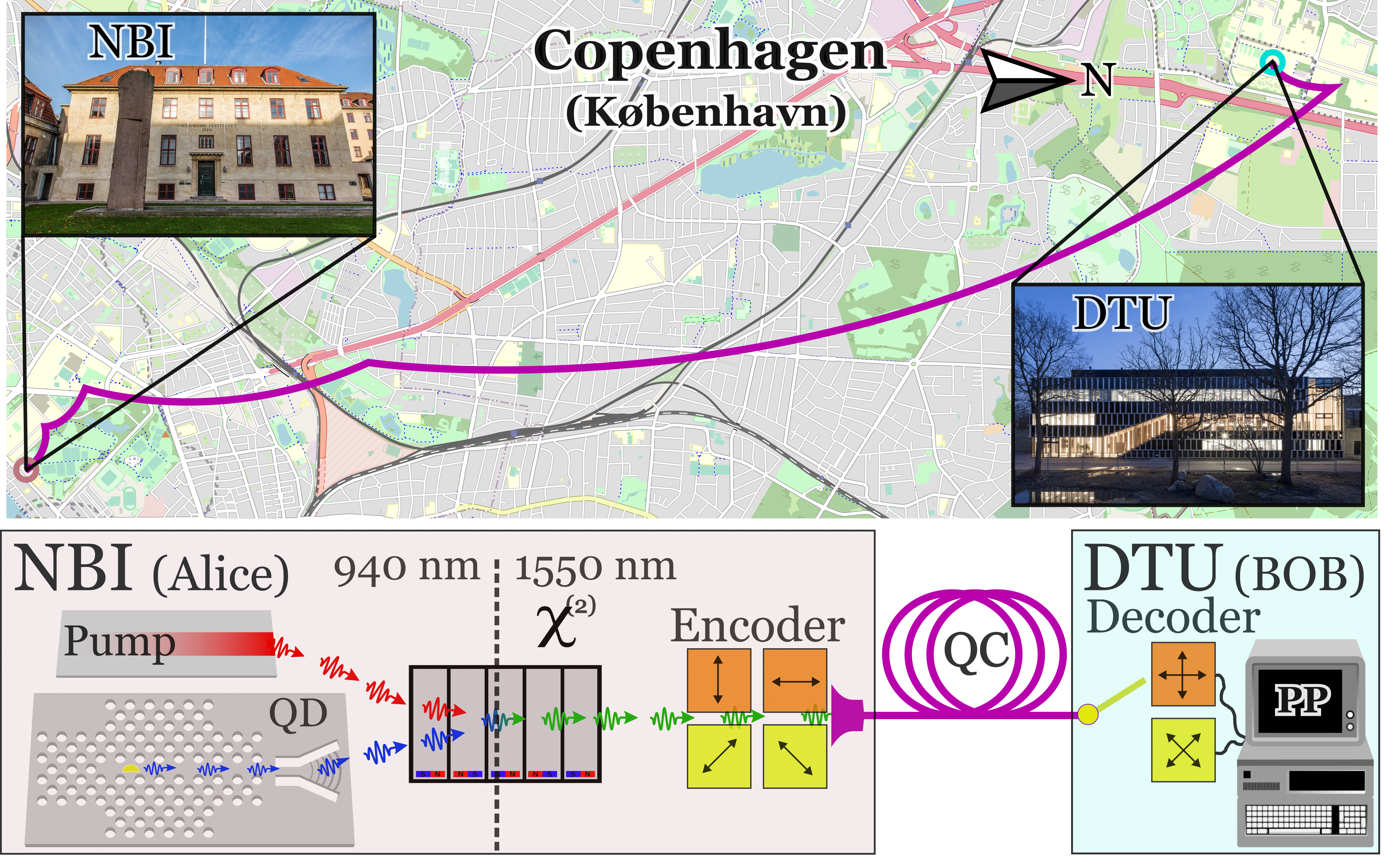}
    \caption{Map showing the quantum channel connecting the Niels Bohr Institute (NBI) and the Technical University of Denmark (DTU), with a length of 18.1 km ( 9.6 dB of channel loss). Bottom panel displays a simplistic schematic of the experiment from the quantum dot (QD) source through nonlinear down conversion to the telecom C-band ($\chi^{(2)}$), QKD encoder, quantum channel (QC) (the physical link), QKD decoder, and post-processing (PP).
    }
    \label{Fig::ChannelMap}
\end{figure}

In contrast, field trials of QKD links based on weak coherent pulses (i.e., attenuated laser) are regularly carried out at the urban-area network level, over dedicated testbeds \cite{PhysRevLett.126.250502,Bacco2019,Ribezzo2022},
or even via satellite \cite{liao2017satellite}. While laser sources, combined with decoy state protocols  \cite{HwangDecoyProtocol} enable remarkably higher communication rates than single-photon sources, the latter offer exciting perspectives in the development of more advanced quantum communication tasks involving entanglement and, eventually, a fully device-independent QKD scheme. Therefore, a field demonstration of a single-photon based QKD link is an important stepping stone in proving the maturity of QDs in a real-world use-case. In fact, single-photon QKD has been so far hindered by the quality of solid-state emitters: 
low or fluctuating photon count rates, emission wavelengths far from the telecommunication bands, and collection setup instability.  

In this work, we employ an advanced single-photon source based on QDs in photonic crystal waveguides \cite{Uppu_2020} and a frequency conversion scheme based on difference-frequency generation (DFG) \cite{Da_Lio_2022} to perform QKD between two districts in the Copenhagen metropolitan area. We use an 18-km-long link made of multiple segments of deployed dark fibre pairs, indicatively shown in Fig. \ref{Fig::ChannelMap}. The fibres connect the sending station (Alice), located at the Niels Bohr Institute in Copenhagen, to the receiving station (Bob) at the Technical University of Denmark in Lyngby. We perform a complete QKD field-trial using single photons with the setup shown schematically in the lower panel of Fig. \ref{Fig::ChannelMap} and analyze the performance of the QD source in terms of security bounds, stability, and actual secret key generation rate.  

\begin{figure*}[t]
    \centering
    \includegraphics[width=1\textwidth]{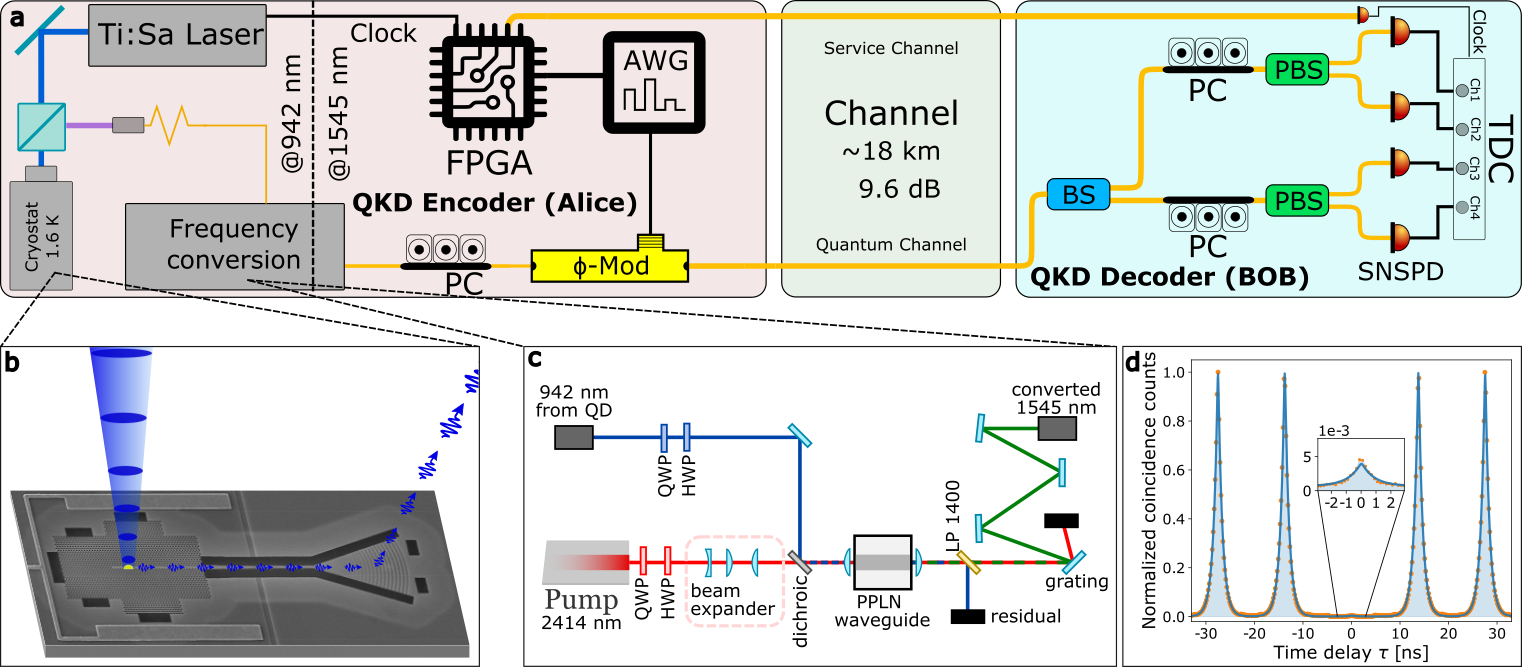}
    Experimental layout of the QKD field-trial demonstrator. \caption{\textbf{Upper panel}: \textbf{a}) Schematic overview of the employed QKD setup with \textcolor[HTML]{BD777F}{Transmitter}, \textcolor[HTML]{84B97F}{Channel} and \textcolor[HTML]{00DFD7}{Receiver}. \textcolor[HTML]{FFBE00}{Yellow lines} mark optical connections (fibres), while \textbf{black lines} are electronic connection (wires). Additionally, the QKD setup consists of a  field programmable gate array (FPGA), an arbitrary waveform generator (AWG), a phase modulator ($\phi$-mod), a beam-splitter (BS), two polarising beam-splitters (PBS), three polarisation controllers (PC), four superconducting nanowire single-photon detectors (SNSPD) and a time-to-digital converter (TDC). \textbf{Lower panel}: \textbf{b}) SEM picture of the nanophotonic structure featuring the QD (yellow annotation), annotated to visualize the single photon source operation with pulsed resonant excitation.  \textbf{c}) Frequency conversion setup including polarisation control (half/quarter-waveplates (HWP/QWP)) and source and pump mode matching lenses. The periodically poled Lithium Niobate (PPLN) waveguide mediates the conversion. A low pass filter (LP) transmitting $>1400$ nm wavelengths is followed by a grating that filters out residual $940$ nm photons and the pump laser. \textbf{d}) Auto-correlation measurement of the down-converted single photons, shown in \textcolor{orange}{orange}, fitted with a double-sided exponential decay convoluted by the instrument response function modeled as a Gaussian distribution in \textcolor{blue}{blue} yielding $g^{(2)}(0)=(0.47\pm0.14)\%$ and a QD lifetime of $(867\pm5)$ ps. Inset shows a zoom-in of the central peak.}
    \label{Fig::Setup}
\end{figure*}

\section{Results}\label{sec::ResultDiscuss}

\subsection{Field trial apparatus}
Fig. \ref{Fig::Setup} \textbf{a} shows the schematic layout of the setup employed in the field trial experiment. The single-photon emitter used in this work is an Indium Arsenide (InAs) QD embedded in a suspended Gallium Arsenide (GaAs) membrane. A photonic crystal waveguide (PCW) is fabricated around the QD allowing near-unity coupling of the QD emission into the waveguide (see Methods for details). The sample is placed in a $1.6$ K closed-cycle cryostat and single photons are collected via a high-efficiency focusing grating coupler using a microscope objective. A single QD transition at 942 nm is excited resonantly with a Ti:Sa mode-locked laser at a repetition rate of 72.6 MHz, resulting in 12 MHz count rate in the fibre (i.e. a source efficiency of $\eta_S=16.5 \%$). A scanning electron microscope (SEM) image of the device is shown in Fig. \ref{Fig::Setup} \textbf{b}.  

To achieve low-loss transmission in fibres, the single photons are down-converted to the telecommunication C-band via difference-frequency generation (DFG) \cite{Da_Lio_2022} in a periodically poled lithium niobate (PPLN) waveguide, as depicted in Fig. \ref{Fig::Setup} \textbf{c}. Using a pump laser at 2414 nm wavelength, the single photons are converted from 942 nm to 1545 nm with a $\simeq 50\%$ end-to-end conversion efficiency. Transport of photons from the single-photon source (located in a different lab) to the frequency conversion setup introduces an additional $29\%$ loss. The down-converted source exhibits a low multi-photon contribution ($g^{(2)}(0)=(0.47\pm0.14)\%$), as verified by auto-correlation measurements performed on the down-converted photons and shown in Fig. \ref{Fig::Setup} \textbf{d}.  A low multi-photon contribution is paramount in order to overcome the threat of PNS-type attacks, and hence a significant parameter in the key distillation process. 

The implemented QKD protocol is the 4-state polarization-based BB84 \cite{BEN84}, where photons are randomly modulated to one of the four polarizations forming the bases $\text{X}=\{|D\rangle, |A\rangle\}$ and $\text{Z}=\{|H\rangle, |V\rangle\}$ in the encoder. This is carried out with a phase modulator actively controlled by an arbitrary waveform generator (AWG). A beamsplitter at the receiver redirects the incoming photons onto either of the two bases ($\text{X}$ or $\text{Z}$), passively at random, upon detection. A master clock phase-locked to the excitation laser by a field programmable gate array (FPGA) is synchronizing the modulation and detection to the photon emission, the latter via an optical service channel parallel to the quantum channel (QC). 
Residual background noise in both the QC and the service channel is observed in the wavelength range 1550-1555 nm, which has been filtered out by tuning the single photon down-conversion process to channel 40 of the International Telecommunication Union-Telecommunication Standardization Sector (ITU-T) and by placing a dense wavelength division multiplexer (DWDM) filter before the state decoder.
An overview of the setup is depicted in Fig. \ref{Fig::Setup} \textbf{a} while the performance is summarized in Table \ref{tab:table11}.

\begin{table}[h]
   \centering
{\begin{tabular}{c|c}
        \textbf{Parameter}  & \textbf{Value}    \\ \hline
        $\nu_{\text{S}}$    & $72.6$ MHz        \\
        $\eta_{\text{QD}}$   & $16.5\%$          \\
        $\eta_\text{T}$     & $71\%$            \\
        $\eta_{\text{FC}}$  & $50\%$            \\ 
        $g^{(2)}(0)$        & $(0.47\pm0.14)\%$ \\
        $\eta_{\text{E}}$   & $55\%$            \\ 
        $\eta_{\text{QC}}$  & $10.9\%$          \\
        $\eta_{\text{R}}$   & $11.4\%$ 
\end{tabular}}
\vspace{0.1cm}
\caption{Overview of the experimental performance including:
 Source rate ($\nu_{\text{S}}$), QD source efficiency ($\eta_{\text{QD}}$), transport efficiency ($\eta_T$) and frequency conversion efficiency ($\eta_{FC}$) efficiencies, second order correlation ($g^{(2)}$), encoder efficiency ($\eta_{\text{E}})$, quantum channel  transmission($\eta_{\text{QC}}$), and receiver efficiency ($\eta_{\text{R}}$) including the detection efficiency.} \label{tab:table11}
\end{table}

\begin{figure*}[t]
   \centering
        \includegraphics[width=\linewidth]{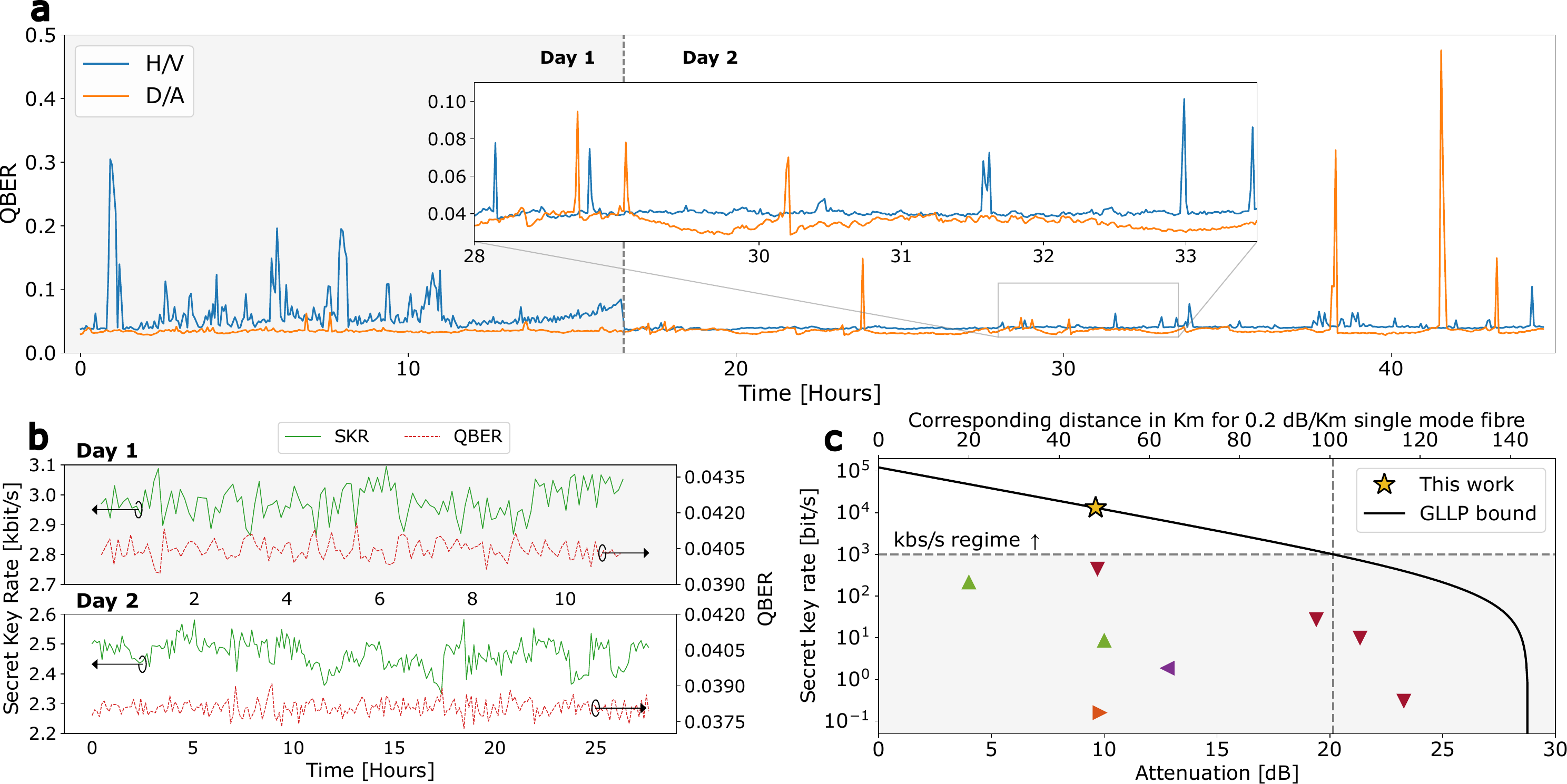}
    \caption{Performance of the QKD field trial. \textbf{a}) QBER for the X and Z basis measurement results during the field trial. Each data point in the figure/inset is the mean of 200/50 seconds, respectively. \textbf{b}) The secret key rate (SKR) and QBER taken during the field trial. The upper and lower figures show the different days of data acquisition, respectively. Each data point represents the mean of 20 post-processing frames.
    \textbf{c}) Comparison of the finite secret key rate of the experimental performance presented in this work based on the asymptotic GLLP bound\cite{GLLPbound}; commonly used for prepare and measure QKD bench marking. The secret key rate is evaluated for the best performance frame with $47.9$ kbit/s raw key and $3.25\%$ QBER and marked with a yellow star. Based on this performance the finite secret key rate is simulated in the asymptotic limit and shown with a black line.
    Triangles mark previous QKD lab-trials with single photon sources at telecom wavelength (\textcolor[HTML]{7E2F8E}{$\blacktriangleleft$}\cite{Intallura_2007}, \textcolor[HTML]{D95319}{$\blacktriangleright$}\cite{ Soujaeff_07}, \textcolor[HTML]{77AC30}{$\blacktriangle$}\cite{ Takemoto_2010}, \textcolor[HTML]{A2142F}{$\blacktriangledown$}\cite{ Takemoto_2015}), all estimated with GLLP-bound.
    }
    \label{fig:Data}
\end{figure*}

\subsection{Field trial data acquisition}
Data was acquired over two different days, referred to as Day 1 and Day 2, over a period of 17 and 30 consecutive hours, respectively. The quantum bit error rate (QBER) measured throughout the whole duration of the field trial is shown in Fig. \ref{fig:Data} \textbf{a}, for the $\text{X}$ and $\text{Z}$ basis, respectively (see Methods). The QBER remains stable throughout the duration of the experiment, especially on Day 2. Short temporary fluctuations in the QBER are observed for the $\text{Z}$ bases on Day 1, which we attribute to the thermal instability of phase modulators in the state encoder in the early hours of the experiment and not to the QD source itself.  
Additionally, sporadic jumps in the value of QBERs are observed over short time intervals for both bases due to overshooting in the active compensation of polarization drifts (see Methods).
In the analysis, we discard all data chunks with very high QBER caused by the polarization drift compensation. 

The secret key is extracted according to a strict finite-size bound \cite{Cai_2009}, with the secret key length given by
\begin{equation}\label{l}
    l_{\text{key}} = \lfloor n\mathcal{A}(1-\text{H}(\Tilde{q}/\mathcal{A})) - \text{leak}_{\text{EC}} - \Delta - \text{leak}_{\text{EV}} \rfloor - \nu_{\text{auth}},
\end{equation}
where $n$ is the number of sifted bits per block used for extracting the secret key bits. We use $n=2\cdot 10^5$ to reduce the impact of finite-size effects while maintaining reasonable computational costs and speed. H is the binary Shannon entropy. The factor $\mathcal{A}=1-p_{\text{m}}/p_{\text{det}}$ corresponds to the leaked information by possible multi-photon emissions \cite{GLLPbound}. $p_\text{m}=g^{(2)}(0)\cdot (\eta_{\text{S}}\eta_\text{E})^2/2$ is the multi-photon probability where $\eta_{\text{S}}$ is the combined QD source, transport, and frequency conversion efficiency, and $\eta_\text{E}$ is the encoder efficiency, while $p_{\text{det}}$ is the detection probability. The upper-bound of the qubit error rate $\Tilde{q}$ is used to account for finite sampling effects together with the finite size correction term $\Delta$ \cite{Cai_2009}.
The terms $\text{leak}_{\text{EC}}$ and $\text{leak}_{\text{EV}}$ correspond to the information leakage during the information reconciliation and error verification, respectively. $\nu_{\text{auth}}$ represents the key portion used for authenticating the classical communication during post-processing.
An adequate description of the key-extraction process is given in the Supplementary Information.

Fig. \ref{fig:Data} \textbf{b} shows the QBER and SKR over the whole duration of data acquisition.
On Day 1, we achieve the highest average SKR of 2.95 kbit/s. Notably, the bound used here allows for the empirical extraction of a usable secret key, accounting for finite-size effects that occur during the implementation of the post-processing. It is therefore significantly tighter compared to the estimates based on the asymptotic GLLP bound \cite{GLLPbound} for single photon sources, which is commonly used to report the SKR. 
While the QBER is higher during Day 1, the final secret key rate is higher compared to Day 2 due to an increased input rate of the unsifted key, see  Table \ref{tab:table:res}. Overall, 361 Mbit of usable secret key has been extracted during the field trial.
 
 \begin{table}[b]
   \centering
\begin{tabular}{c|c|c|c|c|c}
        Day &   QBER & SKR [bits/s] &   KR [kbit/s] & SK [Mbit] & Dur [h]\\
        \hline
        1 & 4.04\% & 2950 & 25.7 & 123 & 11.6\\
        %\hline
        2 & 3.83\% & 2441 & 19.7 & 238 & 27.1
\end{tabular}
\vspace{0.1cm}
\caption{Result overview. Mean final secret key rate (SKR),
mean rate of unsifted key (KR), and total accumulated and usable secret key. Data was collected on two days with the respective running time (Dur).} \label{tab:table:res}
\end{table}%

\begin{table*}[t]
\renewcommand{\arraystretch}{1.2}
\centering
\resizebox{\textwidth}{!}{
\begin{tabular}{c|c|r|l|l|l|l|c|c|c|c|r|l|l}
         Reference & Source          & \multicolumn{1}{|c|}{$\lambda$}& \multicolumn{1}{|c|}{$\nu$} & \multicolumn{1}{|c|}{$\eta_s$} & \multicolumn{1}{|c|}{SPR}  & \multicolumn{1}{|c|}{$g^{(2)}$} & Encoding & Active & Field & Quantum & \multicolumn{1}{|c|}{QBER}  & \multicolumn{1}{|c|}{SKR} & \multicolumn{1}{|c}{$\text{T}_\text{C}$} \\ 
 
        &  & \multicolumn{1}{|c|}{[nm]}& \multicolumn{1}{|c|}{[MHz]}        & \multicolumn{1}{|c|}{[\%]} &\multicolumn{1}{|c|}{[MHz]} & \multicolumn{1}{|c|}{[\%]} &  &  encoding        &trial & channel & \multicolumn{1}{|c|}{[\%]} & \multicolumn{1}{|c|}{[bit/s]} & \\ \hline\hline
              
This Work                 & QD$\downarrow$ & 1545 & \, 72.6  & \, 5.8  & 4.2   & \, 0.47 & Pol.     & \checkmark & \checkmark & Deployed fibre& 3.25    & $13\,200$            & \, 9.6 dB \\ \hline
ref. \cite{Morrison_2022} & QD$\downarrow$ & 1550 & 160.7    & \, 1.09 & 1.75  & \, 3.6  & Pol.     &     -      &    -       & Fibre spool   & $<$2.00 & $94\,800$            & \, 9.9 dB \\ \hline
ref. \cite{Takemoto_2015} & QD             & 1580 & \, 62.5  & \, 5    & 3.13  & \, 0.51 & Time-bin & \checkmark &    -       & Fibre spool   & 2.30    & \, \, \,$450$        & \, 9.7 dB \\ \hline
ref. \cite{Takemoto_2010} & QD             & 1580 & \, 20    & \, 5.8  & 1.16  & \, 5.5  & Phase    & \checkmark &    -       & Fibre spool   & 4.80    & \, \, \,$102$        & 10.0  dB$^*$ \\ \hline
ref. \cite{Soujaeff_07}   & SPDC           & 1550 & \, 12.4  & \, 4.23 & 0.524 & \, -    & Phase    & \checkmark &    -       & Fibre spool   & 4.23    & \, \, \,\, \, $0.16$ & \, 9.8 dB \\ \hline
ref. \cite{Intallura_2007}& QD             & 1300 & \, \, 1  & \, 5.1  & 0.051 & 16.6    & Phase    & \checkmark &    -       & Fibre spool   & 5.90    & \, \, \,\, \, $1.87$ & 12.8 dB   \\ \hline
ref. \cite{Murtaza_2022}  & Molecule       & 766  & \, 80    & \, 8    & 6.4   & \, 2    & Pol.     &   -        &    -       & Fixed att.    & 3.40    & $35\,000$            & \, 9.6 dB \\ \hline
ref. \cite{Collins_2010}  & QD             & 895  & \, 40    & 10      & 4     & 82      & Pol.     & \checkmark &    -       & Fibre spool   & 6.21    & \, \, \,\, $60$      & \, 4.4 dB \\ \hline\hline
ref. \cite{Beveratos_2002}& crystal        & 637  & \, \, 5.3& \, 2.2  & 0.117 & \, 7    & Pol.     & \checkmark &    -       & Free space    & 4.6 \:  & \, $7\,700$          & \: 50 m$^\dagger$ \\ \hline
ref. \cite{Alleaume_2004} & NV Center      & 690  & \, \, 5.3& \, 2.8  & 0.122 & \, -    & Pol.     & \checkmark &    -       & Free space    & 1.7 \:  & \, $1\,600$          & \: 30 m$^\dagger$ \\ \hline
ref. \cite{waks_2002}     & QD             & 930  & \, 76    & \, 0.7  & 0.532 & 14      & Pol.     & \checkmark &    -       & Free space    & 2.4 \:  & $25\,000$            & \: \, 1 m$^\dagger$\\ \bottomrule
\multicolumn{14}{l}{$^{*}$\footnotesize{Based 50 km fibre spool and 0.2dB/km}}\\
\multicolumn{14}{l}{$^{\dagger}$\footnotesize{Attenuation not given, while the free space distance is given here.}} 
\end{tabular}
}
\caption{Comparison to other QKD experiments with single photon sources. QD$\downarrow$ indicates a frequency down-converted QD source. $\lambda$ is source wavelength, $\nu$ is the operation rate of the source, $\eta_s$ the source efficiency and SPR is the single photon rate. Pol. is short for polarisation and att. for attenuator. In order to do a fair comparison to reference experimental achievements the QKD performance parameters, i.e. QBER and SKR, are here compared at $\sim 10$dB channel attenuation where possible, the actual attenuation at comparison is given in the transmission for comparison ($\text{T}_\text{C}$) column. All references are bound against collective attacks using the GLLP bound, except the last three which are only abound against individual attacks.}
\label{tab:Comparison1}
\end{table*}

\section{Discussion}
The $\approx4\%$ QBER measured during the two-day-long data acquisition of the experiment is attributed to two main factors, state preparation of the encoder and channel background noise observed in the C-band. To reduce the background noise, besides placing a DWDM filter, we temporally filtered the incoming photons in a 1 ns temporal filter. Nevertheless, we attribute $\sim1\%$ induced QBER to the signal-to-noise ratio. 
A comparison of the QBER measured in a back-to-back test with weak coherent pulses (WCP) suggests a possible improvement of the QBER by $2-2.5\%$, by improved stabilization of the encoder and the photon transmission line from frequency conversion to the state encoder. 

The secret key extraction is performed under the assumption that no coherence exists between consecutively emitted photons, which is a good approximation when exciting the QD with short resonant $\pi$-pulses \cite{Loredo_2019}. 
The state generation rate can be effectively increased by pumping the QDs at a higher rate. The measured $0.867$ ns lifetime of the QD employed in this work would allow to pump the QD at $\sim0.4$ GHz rate, which would readily correspond to a nearly five-fold increase of the secret key generation rate.
Moreover, Purcell enhancement of spontaneous emission, would open the door to even higher single-photon rates. QD lifetimes below $50$ ps have been reported previously \cite{Tomm_2021}, which allow pumping the QD in the GHz regime. In combination with an optimized low-loss encoder and receiver, a key generation rate approaching the Mbit/s might be achievable on the field-trial channel, showing the great potential of current QD technology for applications that demand a high generation rate.

In Fig. \ref{fig:Data} \textbf{c}, we compare our results to previously reported QKD experiments (all performed in laboratory environment) using deterministic single-photon sources in the C- or O-band. In those experiments, a full QKD protocol has been implemented, which allows for a fair comparison of the asymptotic key rates achievable. Several other single-photon QKD experiments (at different wavelengths or without encoding) have been reported \cite{Collins_2010,Beveratos_2002,Alleaume_2004,Murtaza_2022,Morrison_2022} which cannot be directly compared to ours. A list of reported single-photon QKD experiments is given in the Supplementary information Tab. \ref{tab:Comparison1}. %none of these previous works have implemented a full QKD protocol and extracted secret key rates, therefore a direct comparison with those works is not possible. 
We achieve the highest asymptotic secret key rate reported so far for single-photon QKD, of up to 13.2 kbit/s at a channel loss of 9.6 dB, which for a 0.2 dB/km bare fibre loss would correspond to a distance of 48 km (yellow star in  Fig. \ref{fig:Data} \textbf{c}). Notably, we still achieve kbit/s key rates, which would allow for real-time one time pad encryption of voice recordings \cite{vocoder}, to about 100 km equivalent reach with the commonly used GLLP-bound \cite{GLLPbound}. 

We have presented the first QKD field trial with a near-deterministic single-photon source. We extract actual usable secret key frames and we set the new state-of-the-art of secret key rates in the C-band for single-photon sources to more than 2 kbit/s at $\approx 10$ dB of loss. By demonstrating a stable and usable single-photon connection in a metropolitan environment, we pave the way towards realistic implementations of single-photon based communication infrastrcuture, where deterministic and coherent QD-based single-photon sources constitute a mature resource for advanced applications, including device-independent QKD \cite{DIQKD_Acin,DIQKD_Eva}, one-way quantum repeaters \cite{Borregaard} and, ultimately, the quantum internet \cite{Kimble_2008}.

\section{Methods}
\label{Sec::Method}

\subsection{Single-photon source}
\label{SubSec::SPS}
The single-photon sources are fabricated on a GaAs membrane grown by molecular beam epitaxy on a (100) GaAs substrate. The 180-nm-thick membrane consists of a layer of self-assembled InAs QDs (density \textless 10 $\mu$m$^{-1}$) grown in the middle of an ultra-thin \textit{p-i-n} diode junction, to reduce charge noise and control the charge state while also enabling Stark tuning of the QD emission wavelength. The diode bias is controlled by a stable and low-noise voltage source. Additionally, a distributed Bragg reflector (AlAs/GaAs, 79/66 nm) is grown below the membrane to enhance the vertical collection of photons out of the chip.
The photonic crystal waveguides are fabricated via electron-beam lithography and dry etching following the methods presented in \cite{midolo2015soft}, while electrical contacts to the diode are deposited via electron-beam evaporation and lift-off. 

\subsection{Setup}
\label{Sec::Experiment}

The state encoder is formed by an in-line one-pass polarization modulation \cite{Fadri_InLinePolMod,Jofre_InLinePolMod} in which birefringence of a Titanium-diffused $\text{LiNbO}_3$ crystal induces a relative phase($\phi$) between the two rectilinear components of the input diagonal polarization. The scheme features low loss compared to other modulation techniques \cite{Lucio_Martinez_2009,Agnesi:19} as photons encounter only one insertion loss.

A 4-level RF signal generated by an AWG with levels $\{0, \text{V}_{\pi/2}, \text{V}_{\pi}, \text{V}_{3\pi/2}\}$, corresponding to an induced $\phi=\{0, \pi, \frac{\pi}{2},\frac{3\pi}{2}\}$ relative phase shift, enables the preparation of states in the two bases $\{\text{X}\}$ and $\{\text{Z}\}$. The $\text{X}/\text{Z}$ ratio is decided according to the protocol.

The state encoder and the mode-lock laser are synchronized by first down-sampling the mode-lock laser clock and then triggering the AWG. An FPGA provides a down-sampled clock at 120 kHz which is also transmitted through the service channel for synchronization. Upon triggering, the AWG bursts out a waveform to modulate 605 consecutive photons to one of the four polarization states chosen at random. It should be noted that to guarantee the security of the implementation, the pseudo-random sequence of the states should be replaced with a truly random sequence generated with a quantum random number generator \cite{Zahidy_2021_QRNG}.

The receiver, see Fig. \ref{Fig::Setup}, is comprised of a 50:50 beam-splitter (BS) followed by two sets of polarization controllers and polarizing beam-splitters (PBS). The output port of the PBSs is connected to a superconducting nano-wire single photon detector (SNSPD). Each PBS is aligned to measure one of the bases $\{\text{X}\}$ and $\{\text{Z}\}$ through initial alignment with manual polarization controllers. Two automatic polarization controllers - not depicted in the figure - in line with the manual ones are executing an optimization algorithm based on coordinate-descent search which  compensates for any polarization drift that has occurred in the quantum channel. This maintains the QBER below a certain threshold. The SNSPDs feature ~33 ns of deadtime, 50 Hz dark count, and  $\approx$83\% detection efficiency. The detection events and their accurate timing is registered with a time-to-digital converter (TDC)  with 1 ps resolution for post-processing.

The transmitter and the receiver are connected with a pair of fiber channels of $\approx$18.1 km length, formed by 6 patches, and exhibit 9.6 dB of loss. The fibers transmit the quantum and clock signal. While running the protocol the basis sequence is also transmitted to the receiver for sifting and error estimation.

\section{Supplementary}

\subsection{Comparison to state of the art}
Table \ref{tab:Comparison1} lists the recent QKD experiments which employ single-photon sources based on quantum emitters. For comparison with the state-of-the-art, we have only listed the experiments where a prepare and measure QKD protocol has been reported, e.g. BB84. More advanced protocols, such as the entanglement based E91 protocol \cite{basso2021quantum,Nadlinger_2022} or (measurement) device independent QKD protocols \cite{Zhang_2022,Wei-Pan_300km_QI} are not included, as these currently do not compete in terms of key rate, and the structure of the protocol is fundamentally different. When comparing to other experimental achievements we define the concept of a ``field trial'' as those experiment involving the transmission of the quantum signal over a significant distance exceeding the premises of a single campus area (or the like), on a network infrastructure/channel over which the experimentalists have no control. 

In comparison to the state of the art, we pioneer the art of single-photon QKD over a field-deployed fibre. Moreover, we outperform previously-reported QKD implementations in terms of achieved key rate evaluated with the GLLP bound, with exception of two recent works \cite{Morrison_2022,Murtaza_2022}, which, however, did not implement an active QKD encoding and thus do not account for the losses in the encoding apparatus nor the intrinsic QBER induced by active modulation, which must be anticipated to be higher than that of single state preparation. 

\subsection{Post-processing}
\label{pp-section}
After the signals have been distributed during the physical phase of the QKD protocol, extensive post-processing is required to extract secret key bits. Alice and Bob communicate over a classical public channel and share information about the choice of basis for each bit. Only those cases in which they both chose the same basis by chance will be considered for the following steps, all other measurements are discarded. The remaining, sifted bits are then buffered to fit the required frame size $n=200000$ of the post-processing, while another $m\approx 0.1n$ bits are disclosed to estimate the QBER of each specific frame. Due to bit-flip errors occurring in the quantum channel, there is a mismatch between the sifted keys of Alice and Bob. Low-density parity-check (LDPC) codes \cite{1057683} are used to reconcile the key frames, whereby a single syndrome message is sent from Alice to Bob for each reconciled frame. The correctness of the reconciled keys is verified by exchanging short hashes.

\begin{figure}[h]
   \centering
        \includegraphics[width=\linewidth]{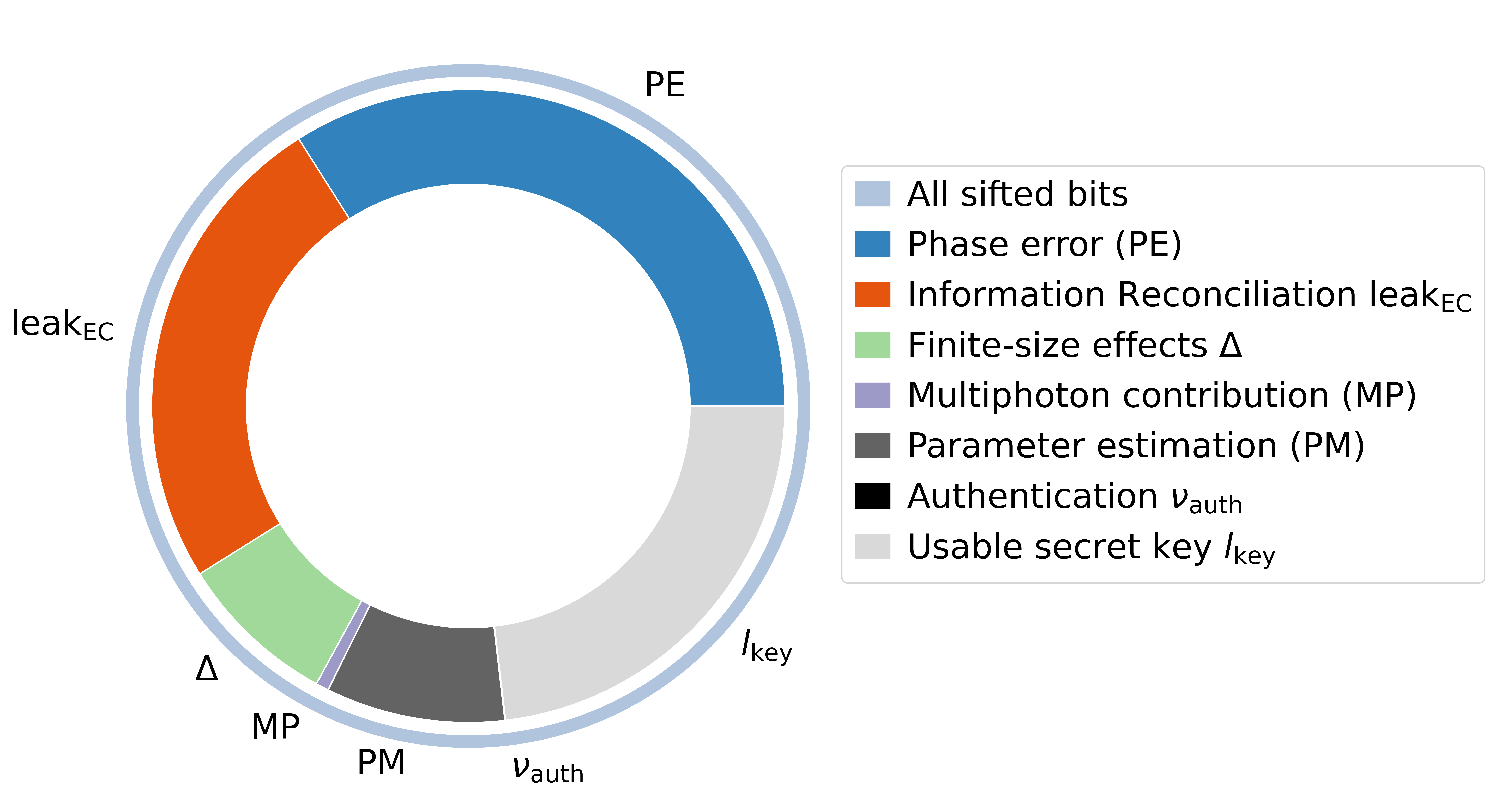}
    \caption{Relative contribution on information leakage of different effects. The leaked information is represented in the number of bits that the secret key has to be shorter than the sifted key due to the respective effects, where the shortening is executed during privacy amplification. The full circle represents all sifted bits of a single frame ($n$), the grey segment shows the size of the final secret key $(l_{\text{key}})$. PE, PM, and MP can all be related to $n\mathcal{A}(1-\text{H}(\Bar{q}/\mathcal{A}))$. Data taken of a single extracted secret key frame.}
    \label{fig:donut}
\end{figure}

Privacy amplification \cite{10.1137/0217014} via the use of Toeplitz matrix hashing \cite{10.1007/3-540-49264-X_24} is then used to extract a shorter, secret key out of the reconciled sifted key. This removes any information leakage. The extracted key is of length $l_{\text{key}}$, given by Eq. \eqref{l}.
Finally, an authentication scheme \cite{9076167} is deployed to ensure the authenticity of the classical channel for each frame. This consumes a small amount of secret key for each frame and requires an initial shared secret key.

Six different LDPC codes with block-size 200000 of code rate 0.65 to 0.9 have been constructed using the PEG algorithm \cite{peg} based on optimized degree distributions \cite{Elkouss_2009}, where the code rate $R$ relates to the ratio of leaked information $(\approx$ syndrome length $m$) to frame size $n$, $R=1-\frac{m}{n}$. We use a rate-adaptive scheme \cite{5650099}.

\begin{table}[h]
   \centering
\begin{tabular}{c|c|c}
        \textbf{Parameter} & \textbf{Symbol} & \textbf{Value} \\
        \hline\hline Source rate&$\nu_{\text{S}}$ &$72.6$ MHz\\
        \hline C-band source efficiency&$\eta_{\text{S}}$  &   $5.8\%$\\
        \hline Second-order correlation &$g^{(2)}(0)$&$(0.47\pm0.14)\%$\\
        \hline Encoder efficiency&$\eta_{\text{E}}$   &   $55\%$\\
        \hline Quantum channel transmission&$\eta_{\text{QC}}$      &   $10.9\%$\\
        \hline Decoder efficiency&$\eta_{\text{Bob}}$     &   $11.4\%$  \\
        \hline\hline Post-processing Blocksize & $n$ & 200000\\
        \hline Bits used for parameter estimation & $m$ & $\approx 0.1 n$\\
        \hline Security Parameter & $\epsilon$      &   $10^{-10}$\\
        \hline Failure prob.  parameter estimation & $\epsilon_{\text{pe}}$   &   $4 \cdot 10^{-12}$\\
        \hline Failure prob.  error verification & $\epsilon_{\text{cor}}$     &   $6 \cdot 10^{-11}$  \\
        \hline Failure prob.  authentication & $\epsilon_{\text{auth}}$ & $10^{-12}$\\
        \hline Detector efficiency & $p_{\text{det}}$  &  $0.83$\\
        \hline Multiphoton emission probability & $p_{\text{m}}$ & $2.5\cdot 10^{-6}$
\end{tabular}
\caption{Overview of the experimental performance and the used parameters for post-processing in the experiment. The security parameters for authentication and error verification correspond to using hashes/tags of 86 and 34 bits, respectively.  The measured efficiency of the information reconciliation is 1.17. The impact of the different terms in Eq. \eqref{l} during the experiment is visualized in Fig. \ref{fig:donut}.} \label{tab:table2}
\end{table}

The security of any extracted key is based on the underlying mathematical security proof of the QKD protocol. Any extracted key is only secure if it adheres to the used proof and if the proof and its assumptions accurately represent the experimental setting. It is common practice to use the asymptotic GLLP \cite{GLLPbound} bound to estimate the secret key rate. However, we directly extract useable and finite size secret key frames. This requires the usage of a tighter security bound \cite{Cai_2009} that respects the finite-size effects that occur during implementation, yielding a key length of:
\begin{equation}%\label{l}
    l_{\text{key}} = \lfloor n\mathcal{A}(1-\text{H}(\Tilde{q}/\mathcal{A})) - \text{leak}_{\text{EC}} - \Delta - \text{leak}_{\text{EV}} \rfloor - \nu_{\text{auth}},
\end{equation}
where $n$ is the number of sifted bits used for extracting the secret key bits. $\text{H}$ is the binary Shannon entropy. The factor $\mathcal{A}=1-p_{\text{m}}/p_{\text{det}}$ corresponds to the leaked information by possible multi-photon emissions \cite{sat}, where $p_{\text{m}}$ and $p_{\text{det}}$ are the multi-photon emission probability and the detection probability of the detector, respectively. $p_{\text{m}}$ can be calculated as $p_{\text{m}}=(\eta_{\text{S}}\eta_{\text{E}})^2g^{(2)}(0)/2$. The upper-bounded qubit error rate $\Tilde{q}$ is used to account for finite sampling effects,

\begin{equation}
    \Tilde{q} = \hat{q} + \frac{1}{2}\sqrt{\frac{2\ln(1/\epsilon_{\text{pe}})+2\ln(m+1)}{m}},%https://www.overleaf.com/project/63725f7fcab8b720eccf0ca2
\end{equation}
where $\hat{q}$ is the maximum likelihood estimate of the QBER, $m$ is the number of bits used for estimation, and $\epsilon_{\text{pe}}$ is the probability that the parameter estimation fails \cite{Cai_2009}. A parameter estimation is considered as failed if $\|q_\infty - \hat{q}\| > \Tilde{q}$, where $q_\infty$ is the estimated value in the limit of infinitely many samples. The finite size correction term $\Delta$ is expressed as \cite{Cai_2009}
\begin{equation}
   \Delta =  7\sqrt{\log_2 (2/\Bar{\epsilon})n} + \log_2(1/\epsilon_{\text{pa}}^2),
\end{equation}
where $\Bar{\epsilon}$ is a parameter that can be optimized together with $\epsilon_{\text{pa}}$ to minimize this contribution. The terms $\text{leak}_{\text{EC}}$ and $\text{leak}_{\text{EV}}$ correspond to the information leakage during the information reconciliation and error verification, respectively. Their value is determined by the length of the syndrome and exchanged hash. The overall security parameter $\epsilon$ can then be written as 
\begin{equation}
    \epsilon \leq \epsilon_{\text{cor}} + \epsilon_{\text{pa}} + \epsilon_{\text{pe}} + \Bar{\epsilon} + \epsilon_{\text{auth}},
\end{equation}
where $\epsilon_{\text{auth}}$ is the security parameter of the authentication. $\epsilon_{\text{cor}}$ is the probability that a frame passes the information reconciliation phase without being corrected, it can be bounded by the collision probability of the used hash function. An overview of the security parameters used as well as measured efficiencies can be seen in Tab. \ref{tab:table2}.  A number of bits $\nu_{\text{auth}}$ of previous secret keys is consumed for authentication each frame, effectively setting the final extracted secret key to $l_{\text{key}}$ in Eq. \eqref{l}, in compliance with the security parameter $\epsilon$. A visualized breakdown of how different effects contribute to information leakage is given in Fig. \ref{fig:donut}.

\section{Acknowledgements}
We acknowledge funding from: The Center of Excellence SPOC (ref DNRF123), Innovations fonden project Fire-Q (No. 9090-00031B), Danish National Research Foundation (Center of Excellence Hy-Q DNRF139), and Styrelsen for Forskning og Innovation (FI) (5072-00016B QUANTECH). L.M. acknowledges the European Research Council (ERC) under the European Union's Horizon 2020 research and innovation programme (Grant Agreement No. 949043, NANOMEQ). D. B. acknowledges the programme Rita Levi Montalcini (PGR19GKW5T).

%\section{Author contributions}

%\section{Data availability}


\bibliography{sn-bibliography}% common bib file

\end{document}